\documentclass[pre,twocolumn,a4paper,floatfix]{revtex4}
\usepackage{graphicx}
\usepackage{amssymb}
\usepackage{dcolumn}

\begin{document}

\title{Citation Networks in High Energy Physics}
\date{\today}

\hyphenation{data-base clear-ly SPI-RES de-scrib-ed}

\author{S. Lehmann}
\author{B. Lautrup}
\author{A. D. Jackson}
\affiliation{The Niels Bohr Institute, Blegdamsvej 17, DK-2100
Copenhagen {\O}, Denmark}

\begin{abstract}
The citation network constituted by the SPIRES data base is
investigated empirically. The probability that a given paper in
the SPIRES data base has $k$ citations is well described by simple
power laws, $P(k) \propto k^{-\alpha}$, with $\alpha \approx 1.2$
for $k$ less than 50 citations and $\alpha \approx 2.3$ for 50 or
more citations.   Two models are presented that both represent the
data well, one which generates power laws and one which generates
a stretched exponential. It is not possible to discriminate
between these models on the present empirical basis.  A
consideration of citation distribution by subfield shows that the
citation patterns of high energy physics form a remarkably
homogeneous network. Further, we utilize the knowledge of the
citation distributions to demonstrate the extreme improbability
that the citation records of selected individuals and institutions
have been obtained by a random draw on the resulting distribution.
\end{abstract}

\maketitle

\section{Introduction}

Recently, the study of networks has become a part of statistical
physics. This connection between sociology, where social networks
have been studied since the late sixties \cite{milgram:67}, and
statistical physics, has arisen because the methods of statistical
physics have proven to be valuable tools when analyzing a variety
of complex systems; amongst these are complex networks
\cite{albert:99,barthelemy:99,newman:99b,menezes:00,newman:00d,%
barabasi:99a,cohen:00,%
callaway:00,moore:00,newman:00c,barrat:00,moukarzel:99,%
dorogovtsev:00,krapivsky:00a,dorogovtsev:00a,kulkarni:00,%
kleinberg:00,watts:98,barabasi:99,kumar:00}. The real world
networks that have been studied by physicists include the world
wide web \cite{albert:99,kumar:99,broder:99}, the internet (the
physical connections between computers)
\cite{faloutsos:99,yook:01,pastor:01}, email networks
\cite{newman:02}, phone call networks \cite{aiello:01}, movie
actor collaboration networks \cite{newman:00c,barabasi:99},
metabolic networks \cite{jeong:00}, the power grid of the united
states \cite{watts:98}, and numerous others. Closer to the subject
of the network of citations, the properties of scientific
co-author networks have been studied in
\cite{newman:01,newman:01a} and modelled in \cite{barabasi:02}.

The present paper focuses on the topology of the network of
citations of scientific publications. In this network every paper
is a node, and an edge (i.e.~a link between two nodes) arises when
one paper is cited by another. Clearly, this is a directed
network, that is, every edge has a direction; normally a reference
from one paper to another actually rules out a reference in the
other direction (reciprocity $\approx 0$). The data presented in
this paper is the number of citations accumulated by each paper;
we do not have access to the list of references for each paper.
Therefore, we will mainly be concerned with the in-bound degree
(citation) distribution of papers in the SPIRES data base.

In addition to the pure theoretical interest in complex networks,
the subject matter of this paper should be of interest to
physicists for a completely different reason. It has been
recognized since the early 1970s that citations can provide a
quantitative measure of scientific excellence \cite{isi:02}. Many
studies (e.g., \cite{adam:02} and references therein) have shown
that this tool must be used with considerable care. Different
scientific environments have different publishing and citation
habits, and these differences must be reconciled before
comparisons can be made across field boundaries. Nevertheless,
citation studies have become a standard measure for the evaluation
of journal impact or of the quality of university departments.
Just as a study of email networks can enlighten us about the
spread of computer viruses, and a study of the structure of the
internet can be used to estimate the amount of damage caused by
router breakdown, the study of citation networks can help us
understand and quantify scientific excellence.

\subsection*{Past investigations}

Given the level of interest in complex networks and citation data,
surprisingly few serious studies of citation networks have been
performed by physicists. In 1957 Shockley \cite{shockley:57}
argued that the publication rate for the scientific staff at
Brookhaven National Laboratory was described by a log-normal
distribution. In 1998, Laherrere and Sornette \cite{Laherrere:98}
suggested that the number of authors with $k$ total citations,
$N(k)$, of the 1120 top-cited physicists from 1981 to 1997 is
described by a stretched exponential ($N(k)\propto
\exp[-(k/k_0)^\beta]$, $\beta \approx 0.3$).  Note however that
this study focuses on the total number of citations of topcited
authors and not on the distribution of citations of publication as
is the case in the present paper. Also in 1998, Redner
\cite{redner:98} considered data on papers published in 1981 in
journals catalogued by the ISI as well as data from Phys.\,Rev.\,D
vols.\,11-50 and concluded that the large-$k$ citation
distribution is described by a power-law such that $N(k)\propto
k^{-\alpha}$ with $\alpha \approx 3$.

In the present paper, the statistical material is of a much higher
quality than in the papers mentioned above; we present the results
of a study of the SLAC SPIRES data base
\footnote{\texttt{http://www.slac.stanford.edu/spires/hep}}. The
ISI data set studied in \cite{redner:98} is materially larger
(783,339 papers) than SPIRES data set.  However, the ISI data used
by Redner contains papers published in a single year in a variety
of scientific disciplines (including medicine, biology, chemistry,
physics, etc.).  There are neither a priori arguments nor data to
indicate that citation patterns in these fields are sufficiently
uniform to justify their treatment as a single data set.  The
SPIRES hep data is collected from a well-defined area within
physics, i.e.\ high energy physics, and has been accumulated
systematically by the SLAC library since 1962 \cite{oconnell:02}.

To be specific, the data used below was retrieved from the SPIRES
mirror at Durham University on August 14, 2002. We will henceforth
refer to this as the SPIRES data base.  Since the SPIRES data base
is dedicated to papers in high energy physics, it is natural to
assume that it is relatively homogeneous.  One of the purposes of
the present work is to determine the extent to which citation
patterns in the categories of theory, phenomenology, experiment,
instrumentation, and reviews are in fact comparable.  We will then
present the citation probability for the SPIRES data base.

\section{The citation distribution}
\subsection*{Basic statistics}

The SPIRES data base contains 501,531 papers. Of these papers
there are 196,432 non-journal papers (e.g.\ preprints and
conference proceedings) for which citation information is not
available.  A fraction of the remaining papers seem to have been
removed from the data base.  In other cases, subfield designations
are not available.  Thus, we have restricted our attention in the
following to the network of 281,717 papers (i.e.\ roughly 56\% of
the SPIRES data base) for which both citation information and
subfield designations are available. Table \ref{tab:probs} shows
the probability, $P(k)$, of a SPIRES paper having $k$ citations
for $0 \le k \le 4$.  An ``atomic'' histogram of the full citation
data is shown in Figure 1.
\begin{table}[tb]
\begin{center}
\begin{ruledtabular}
\begin{tabular}{lccccc}
{}              & $P(0)$ & $P(1)$ & $P(2)$ & $P(3)$ & $P(4)$ \\
{}              & {}     & {}     & {}     & {}     & {}     \\
\hline
Theory          & 0.2884 & 0.1226 & 0.0815 & 0.0590 & 0.0472 \\
Phenomenology   & 0.2150 & 0.1103 & 0.0762 & 0.0618 & 0.0488 \\
Experiment      & 0.2677 & 0.1023 & 0.0704 & 0.0518 & 0.0441 \\
Instrumentation & 0.6169 & 0.1206 & 0.0622 & 0.0385 & 0.0267 \\
Review Articles & 0.2167 & 0.1038 & 0.0670 & 0.0496 & 0.0403 \\
{}              & {}     & {}     & {}     & {}     & {}     \\
Total           & 0.2901 & 0.1171 & 0.0775 & 0.0574 & 0.0458 \\
\end{tabular}
\end{ruledtabular}
\end{center}
\caption{The probability of a paper in the SPIRES data base having
$k$ citations for $0 \le k \le 4$ as a function of subfield. The
total number of papers in each subfield is: 159,946 (theory),
68,549 (phenomenology), 28,527 (experiment), 19,637
(instrumentation), and 5,058 (review articles).  The ``total''
data entries are obtained directly from the subfield data.  The
total number of papers in the data set is
281,717.}\label{tab:probs}
\end{table}

One of the most striking features of this data set is the large
number of papers (some 29\%) which are uncited. Note that we have
not applied any correction for self-citation. The removal of
self-citations would make the fraction of uncited papers
materially higher.  In the same vein, 74\% of the papers in our
network have 10 or less citations.  In contrast, 6.2\% of the
papers have 50 citations or more and only 131 papers ($\approx
0.05\%$) are cited 1000 times or more.  The mean number of
citations in this sample is 14.6, which is considerably larger
than the median of 2.3 citations, implying that a paper with the
average number of citations is substantially more cited than the
``average'' paper. The large factor between mean and median
citations suggests that the citation distribution has a very long
tail with a small fraction of highly cited papers accounting for a
significant fraction of all citations.  This is indeed the case.
Approximately 50\% of all citations are generated by the top 4\%
of the all papers; the lowest 50\% of papers generates only 2\% of
all citations. The rates of citation production by these two parts
of the data set differ by a factor of approximately 310. These
observations regarding citations in SPIRES suggest that the
citation distribution follows a power law. As we shall see, this
is qualitatively correct.

\begin{figure}[tb]
\begin{center}
\includegraphics[width = \hsize]{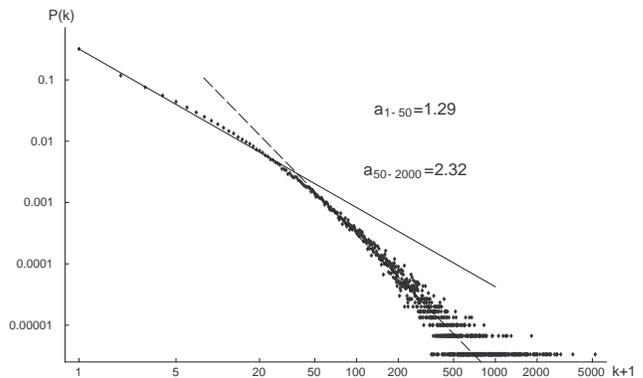}
\end{center}
\caption{An ``atomic'' histogram of the citation distribution of
the total data set showing the normalized probability, $P(k)$,
that a paper has $k$ citations, here plotted versus $k+1$. The
straight lines in the low and high citation regimes have slopes
$-1.29$ and $-2.32$, respectively. Note the logarithmic scales.}
  \label{fig:totalrawplot}
\end{figure}

Figure \ref{fig:totalrawplot} shows a log-log representation of
the distribution of citations in the SLAC SPIRES data base.  The
data suggest that this citation distribution is remarkably well
described by two power laws.  The distribution, $P(k)$, is
approximately proportional to $(k+1)^{-1.3}$ for $0 \le k \le 49$
and to $(k+1)^{-2.3}$ for $k \ge 50$.  Before turning to a more
quantitative description, we consider the homogeneity of the
SPIRES data.

\subsection*{Homogeneity of the data base}

Even though the SPIRES data base is devoted exclusively to papers
in high energy physics, it is relatively easy to imagine
mechanisms which could lead to different citation patterns and
thus different network topologies in the five different subfields
into which the SPIRES data base is divided; these fields are
theory, experiment, phenomenology, reviews, and instrumentation.
Experiments in high energy physics are expensive and manpower
intensive.  Program committee approval is tantamount to a
pre-review of the work.  The number of co-authors is large.  Under
such conditions, it might be reasonable to expect rather fewer
minimally cited papers.  By contrast, the number of co-authors of
papers in the theory and phenomenology sections of SPIRES is far
smaller, and the relatively low cost of such work permits the
production of papers which might not survive pre-reviewing.  In
short, theory and phenomenology subfields might have a larger
probability for minimal citation.  Similarly, one could argue that
review papers, which are often ``commissioned'' by journals and
frequently written by recognized experts, might enjoy higher
citation rates---just as one could conceive of mechanisms such
that the instrumentation subfield might include more minimally
cited papers.  With such a priori expectations, it is of obvious
importance to determine citation distributions separately for each
subfield.  Fortunately, SPIRES is well-suited for such a study.

Some indications of the differences between the five categories
can be seen from Table \ref{tab:probs}.  The probability of having
$\le 4$ citations is $59.9$, $53.6$, $51.2$, $47.7$, and $86.5$\%
for theory, experiment, phenomenology, reviews, and
instrumentation, respectively.  While the fraction of minimally
cited review papers is clearly smaller than that for the full data
set, this effect is not dramatic.  Instrumentation papers,
however, stand out.  The probability that an instrumentation paper
will receive $\ge 5$ citations is almost $3$ times smaller than
that for the full data. The differences between citation
probabilities in theory, experiment and phenomenology are
surprisingly small.  These trends are supported by the full data
set.  We find, for example, that only 146 of the 19,637
instrumentation papers ($\approx 0.7\%$) have 50 or more
citations.  This is to be compared with 6.2\% for the full data
set.  By contrast, approximately $14$\% of review papers have $\ge
50$ citations.  The $3$\% of review papers with $\ge 1000$
citations is significantly larger than the probability of $0.05$\%
for the complete data set.  In short, instrumentation and review
papers, which account for some 9\% of the full data set, clearly
follow different citation distributions.  This can reflect a
different underlying dynamical picture for citations in these
categories; it can also be an indication that review papers have a
higher average quality and instrumentation papers a lower.
Whatever the explanation, we choose to exclude these two small
categories from further consideration.  Any decision to use
citation data as a measure of scientific ``quality'' should not be
made so lightly. Ultimately, however, it must be based on a
subjective evaluation of the relative quality and importance of
papers published in the various categories.

\begin{figure}[tb]
\begin{center}
\includegraphics[width = 1\hsize]{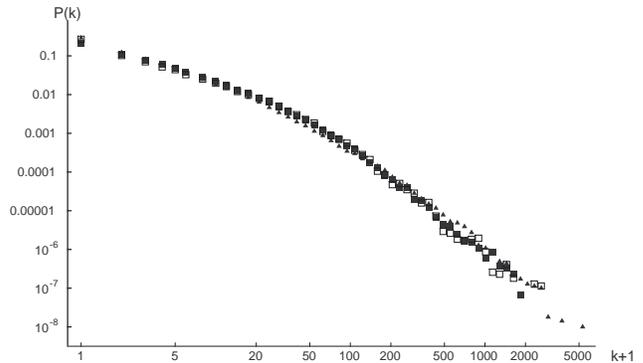}
\end{center}
\caption{Citation distributions for the categories theory
($\blacktriangle$), phenomenology ($\blacksquare$), and experiment
($\square$).} \label{fig:comparison}
\end{figure}

The homogeneity of citation patterns in the categories of theory,
experiment and phenomenology is supported by the binned histograms
show in Figure 2.  Given the logarithmic scale of this figure, the
three citation distributions are essentially indistinguishable
over the full range of $0$ to $5000$ citations.  This agreement is
remarkable in view of the fact that it persists over almost seven
orders of magnitude.  Phenomenology and experiment are in the best
agreement with a maximum discrepancy of some $15$\% found in the
vicinity of $k = 50$.  The maximum discrepancy of approximately
$50$\% between theory and the other two categories is also found
in the vicinity of $k=50$ with materially smaller discrepancies
for other values of $k$.  It would be valuable to know if these
differences are ``statistically significant''.  To this end, it is
tempting to assign errors in each bin proportional to the square
root of the number of papers in each bin and perform a $\chi^2$
fit. This temptation should be resisted.  The assumption required
for such an exercise to be meaningful is, of course, that the data
in the various bins is statistically independent.  This
assumption, which can be demonstrated to be false, is in evident
contradiction with our reason for studying citation distributions
in the first place: We believe that there is a positive
correlation between the intrinsic quality of a scientific paper
and the number of citations which it receives, and we also believe
that ``good'' papers are produced by ``good'' scientists. The
consistency of these three data sets is, however, sufficient for
many applications.  In the following, we will work with this final
data set of 257,022 articles.  The resulting distribution is shown
in Figure 3.
\begin{figure}[tb]
\begin{center}
\includegraphics[width=\hsize]{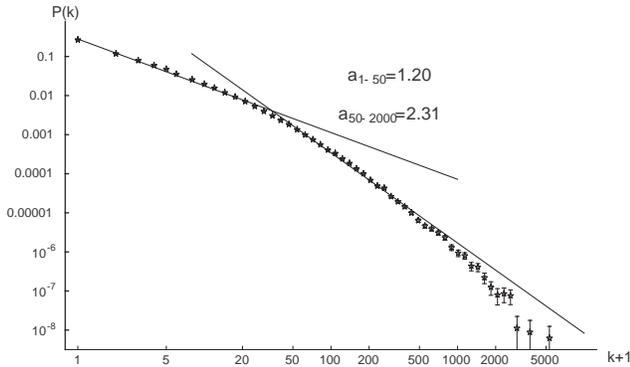}
\caption{A binned histogram of the total data set without review
and instrumentation papers.} \label{fig:finaldatabin}
\end{center}
\end{figure}

There is another and quite different potential source of
inhomogeneity in the SPIRES data base.  The distribution of the
number of authors who have written $y$ papers is a monotonically
decreasing function of $y$. Approximately $91$\% of the individual
authors in the theory data set have written a total of less than
20 papers. Presumably, this effect is due to the large number of
young physicists who leave academic physics either immediately
following their PhD or relatively soon after. Thus, we have also
considered citation probabilities for papers collected
author-by-author. The reason we have solely considered the theory
subset is that the author-by-author data unavoidably weighs papers
by the number of coauthors. As we have noted earlier the theory
subset has fewer authors per paper (typically 1-3) than for
instance the experiment subset where some papers have as many as
1500 authors. For the theory data, the resulting distribution is
similar to that of Figure 3, but not identical. The virtue of such
an author-by-author approach is that it allows us to exclude
authors on the basis of the total number of papers they have
produced. For example, we have compared the citation distributions
of papers by all authors with that of papers written (or
co-written) by authors with more than 20 total papers.  The
differences are extremely small (i.e., similar to those seen in
Figure 2) and again indicated the striking homogeneity of the
SPIRES data base.

\subsection*{The form of the distribution}

Having established the homogeneity of the bulk of the data base or
equivalently the homogeneity of sub-network topologies, we now
turn to a closer look at the form of the distribution. It is clear
from the figures that the distribution cannot be described by a
single power law over the entire range of citations. It is,
however, approximated well by two independent power laws in the
low ($k \le 50$) and high ($k \ge 50$) domains. Thus, $P(k)
\approx (1+k)^{-\alpha}$ in each region with $\alpha_< = 1.20$ and
$\alpha_> = 2.31$.  If we insist on a relative normalization such
that the two forms are equal at $k = 50$ and chose the global
normalization to ensure that the total probability is $1$, the
data is reproduced with surprising accuracy.

We believe that these different power laws probably reflect
differences in the underlying dynamics of citations in the high
and low citation regions.  That different dynamics rule the two
regimes seems clear. The bulk of the papers in the minimally cited
part of the distribution are ``dead'' in the sense that they have
not been cited within the last year or more (and will probably
never be cited again).  Of course, this part of the distribution
also contains vigorous young papers of high quality whose citation
count is increasing.  However, dead papers vastly outnumber the
live population. In the highly cited region, virtually all papers
are still alive, with even the oldest of them acquiring new
citations regularly.  It seems highly likely that citation
patterns for such papers are quite different from those of
minimally cited papers that are most often cited only by the
author and close colleagues.  Further considerations regarding the
temporal evolution of citation networks can be found in
\cite{redner:98} and for the SPIRES hep data base in particular,
in a forthcoming paper by the present authors.

\subsection*{The asymptotic tail}

\begin{figure}[tb]
\begin{center}
\includegraphics[width=\hsize]{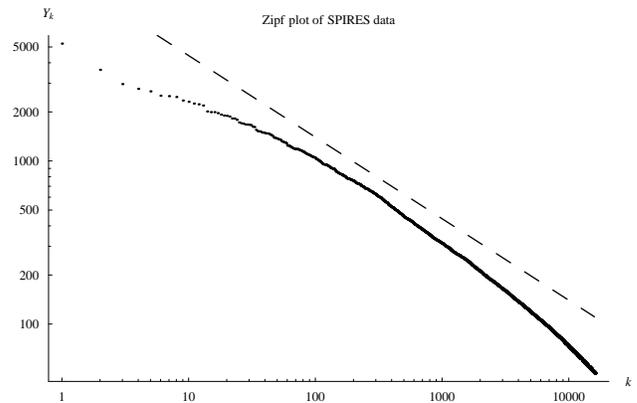}
\caption{A Zipf plot of the citation distribution. For visual
reference a line of slope $-\frac{1}{2}$, corresponding to $\alpha
= 3$, is also plotted.} \label{fig:zipfplot}
\end{center}
\end{figure}

We now consider the large-$k$ tail of the distribution.  Data is
too sparse for a direct analysis in the region of 2000--5000
citations.  Thus, in \cite{redner:98} a Zipf plot is used to
highlight this section of the distribution. A Zipf plot is a plot
of the $n$th ranked paper versus the number of citations of this
paper, $Y_n$. (The most cited paper is assigned rank $1$.)  The
intuitive reason why the Zipf plot is well suited for analyzing
the large-$x$ data is that it provides much higher resolution of
the high citation end of the distribution.  On a doubly
logarithmic scale, the high citation data is placed at the
beginning and is not as compressed as in the plots of $N(k)$
versus $k$ shown in Figures 1--3. Figure \ref{fig:zipfplot} is a
Zipf plot of the final data set.

In reference \cite{redner:98} a similar Zipf plot is used to argue
that the large-$k$ tail of the ISI citation distribution for
scientific papers appeared to be governed by a power law, $1/k^3$.
This is not the case for the SPIRES data. Indeed, Figure
\ref{fig:zipfplot} indicates that the large-$k$ tail of this
highly homogeneous data set is not described by any asymptotic
power-law.  The same conclusion can be drawn from Figure 3, where
a simple power law in the high citation region tracks the data
accurately through four decades until the data begins to cut off.
Although the high-$k$ data is sparse, one can present more
quantitative indications of this cut off.  If the power-law seen
in figure \ref{fig:finaldatabin} applied for arbitrarily large
$k$, as proposed by \cite{redner:98}, we would expect to find 33
papers with more citations than the maximum 5242 citations
actually found in the data set.  The most cited of these papers
should have approximately 55,000 citations. Assuming an
asymptotic power-law, the probability of drawing 257,022 papers at
random with no paper having more than 5242 citations is
approximately $10^{-14}$.

There is a simple explanation for the large-$k$ data which seems
reasonable for a data set like SPIRES, which contains a
significant number of truly important papers. Papers of high
quality and lasting importance can literally be `canonized' and
pass into the received wisdom of physics which no longer requires
citation. Many theoretical physicists publish about `Goldstone
bosons', but few feel the need to cite the original papers.
Indeed, the careful reader will stop to think what special point
is being made when Einstein is cited on special
relativity\cite{einstein:05}. Since only ``mortals'' are cited,
the power law must end. In the absence of such a cut off,
reference \cite{einstein:05} should have been cited by 20\% of the
papers in SPIRES.  This does not seem unreasonable.

\subsection*{Ambiguity of representation}
Because of the cut-off for the high-citation data, there is a
certain ambiguity in determining which mathematical representation
should be chosen for the citation distribution. This ambiguity can
be illustrated by an example. We have modelled the citation
distribution using modifications of the scale free model proposed
by Barab{\'a}si and Albert \cite{barabasi:99}. Model $A$ starts
out with $m_0$ papers with one citation (one incoming edge). At
each time step a paper is added that has one citation and $m$
references (outbound edges). Each of these references link to a
paper $i$ already in the data base with probability $\Pi_A(k_i)$,
proportional to the number of inbound edges $k_i$ of node $i$,
raised to some power $\eta$, that is, $\Pi_A(k_i)\sim k_i^\eta$.

\begin{figure}[htbf]
  \includegraphics[width=\hsize]{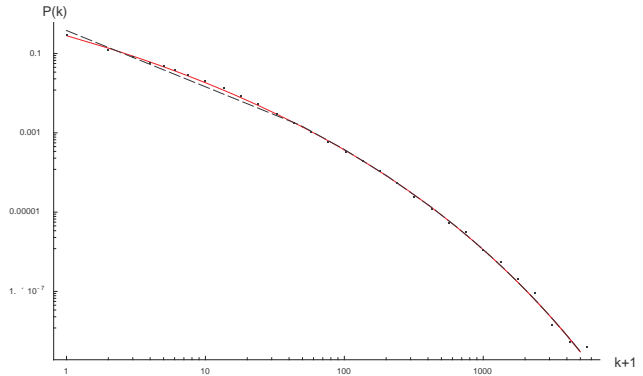}
  \caption{Comparing model $A$ and data. The analytical solution of
  the citation model (solid line) and normalized data from the
  theory subfield (data points). The dashed line is the functional
  approximation (equation (\ref{eq:sublinearfinal})). The parameters used for the model are $m=14.5$, which
  corresponds to the mean number of citations in the theory subfield
  and $\eta=3/4$.}
  \label{fig:solcomparison}
\end{figure}

To solve model $A$ analytically, one can for instance use the rate
equation approach proposed in \cite{krapivsky:00a}. The solution
that is relevant for our data is valid in the regime $0 < \eta <
1$ and in the limit of many time steps; solving the rate equation
under these constraints yields the in-degree distribution
\begin{equation}\label{eq:sublinearprod}
P_A(k)=\frac{\mu}{m}k^{-\eta}\prod_{j=1}^{k}
        \left(\frac{\mu}{mj^\eta}+1\right)^{-1},
\end{equation}
where $\mu(\eta)$ is defined (implicitly) by
$\mu=\sum_{k\ge1}k^\eta P(k)$. This probability is well
approximated by
\begin{equation}\label{eq:sublinearfinal}
    P_A(k)\approx
    \frac{\mu}{\mu+m}k^{-\eta}\exp\left\{-\frac{\mu}{m}\frac{k^{1-\eta}-2^{1-\eta}}{1-\eta}\right\}.
\end{equation}
In Figure \ref{fig:solcomparison}, we have plotted the binned data
from the theory subset along with the exact solution
(Eq.~(\ref{eq:sublinearprod}); solid line) and the approximation
(Eq.~(\ref{eq:sublinearfinal}); dashed line). The fit is
excellent.

Now, let us look at another variation of the model from before,
model $B$. In this version, each paper comes with $w$ ``ghost
citations'' and $m$ references as before; we set $\eta = 1$, so
that $\Pi_B(k_i)\sim k_i+w$. Proceeding as in the case above,
model $B$ can be solved to yield (with the ghost citations
subtracted)
\begin{equation}\label{eq:modelbexact}
    P_B(k)=
    \frac{(m+w)\Gamma(3+w+\frac{w}{m})\Gamma(w+k)}{(1+m+w+2mw)\Gamma(w+1)\Gamma(2+w+\frac{w}{m}+k)}
\end{equation}
for the citation distribution in model $B$.

\begin{figure}[htbf]
  \includegraphics[width=\hsize]{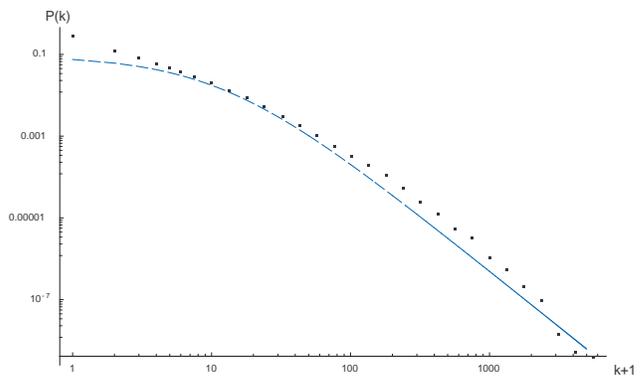}
  \caption{Comparing model $B$ and data. Again the data from the theory
  subsection is represented using dots, whereas the dashed line is
  given by equation (\ref{eq:modelbexact}). The values of
  $m$ and $w$ are set to 15 and 9, respectively, this corresponds to an
  asymptotic power law with slope $\gamma_B=2.6$}
  \label{fig:solcomparison2}
\end{figure}
The probability $P_B(k)$ is an asymptotic power law; in the limit
$k\gg 1$, we have that $P_B(k)\sim k^{-\gamma_B}$, where $\gamma_B
= 2+(w/m)$. The fit to the data is not as compelling as for model
$A$, but it precisely illustrates the ambiguity in deciding on how
to represent the data. We have two representations of the data
with \emph{very} different mathematical properties (the stretched
exponential and the asymptotic power law). Within the range of
$k$'s available before the cut-off sets in, it is difficult
(quantitatively) to discern the power law from the stretched
exponential representation when comparing with the
data---especially so on a log-log scale. In the highly cited
regime, where the exponential begins to dominate Equation
(\ref{eq:sublinearfinal}), and the differences of the two
representations begin to manifest themselves, the presence of the
cut-off makes us unable to draw any conclusions on which
representation to choose, as is amply underlined by Figures
\ref{fig:solcomparison} and \ref{fig:solcomparison2}.

We believe that the mechanisms behind the cut-off are real, but on
the basis of the data available to us at the moment, it is
impossible to estimate its impact on the citation distribution. In
the same vein, we find it probable that the two power laws reflect
different dynamics in the high and low citation regime, but as it
is reflected in the minimal models described above, it is of
course also possible to take a different stand and claim that the
distribution of citations has stretched exponential nature. Using
arguments similar to those of the last section, drawing on the
probability distribution defined by Equation
(\ref{eq:sublinearfinal}), we would expect to find a little less
than 1 paper with more than $5,242$ citations, if this
distribution applied to arbitrarily large $k$; with a data set of
$159,946$ papers, we would expect the maximally cited paper to
have about 4,700 citations. Again, this fidelity to the data is
alluring, but with the data available to us at the moment it is
impossible to draw decisive conclusions either way.

This conundrum has been frequently encountered in the literature.
In the case of distributions of citations, ref.
\cite{Laherrere:98} found the distribution of citations of
scientists to be a stretched exponential, whereas it was argued in
\cite{redner:98} that the citation distribution of papers was
described by an asymptotic power law. The same data was attempted
fitted to a curve $\sim (k_i + {\rm const})^{-\alpha}$ in a later
paper \cite{tsallis:99}. As demonstrated above, our data is of a
much higher quality than the ISI and PRD data sets discussed in
these two papers, but it seems to be the case that even with
access to the highly homogeneous SPIRES data base, the cut-off
mechanism still leaves room for speculation as to the topology of
the citation distribution. Arguments regarding the ``microscopic''
citation mechanisms will have to be made before any model of the
citation network based on the data presently available can be
taken seriously.

Proceeding to a more general arena, the very same problem also
appears in other complex networks. For instance, Newman describes
the distribution of the number of collaborators per publication in
different data bases (amongst these, SPIRES) as a stretched
exponential \cite{newman:01c}, but having acquired more
statistical material, the very same distribution is tentatively
described as two power laws \cite{newman:01} (after inspiration
from \cite{barabasi:02}). In conclusion: For the range of $k$'s
available to us, both the two-power-law structure and the
stretched exponential are reasonable fits to the data.

\section{An application}

Having determined the form of the distribution of the SPIRES data
base and demonstrated its homogeneity, it is interesting to show
that it can be put to practical use.  Here, we present one such
application. The ``citation summary'' option in the SPIRES data
base returns the number of papers for a given author with
citations in each of six intervals.  These intervals and the
probabilities revealed by our analysis that papers will fall in
these bins are given in Table \ref{tab:citsum}. The probability,
$P$, that an author's actual citation record of $M$ papers was
obtained from a random draw on the citation distribution is
readily calculated by multiplying the probabilities of drawing the
author's number of citations in the different categories, $m_i$,
and correcting for the number of permutations.
\begin{displaymath}
P = M! \, \prod_i \, \frac{p_i^{m_i}}{m_i !} \ .
\end{displaymath}
\begin{table}[tb]
\begin{ruledtabular}
\begin{tabular}{lrd}
Paper category    & Citations & \multicolumn{1}{r}{Probability}\\
  {}              & {}        &       \\ \hline
Unknown papers    & 0         & 0.267   \\
Less known papers & 1--9      & 0.444   \\
Known papers      & 10--49    & 0.224   \\
Well-known papers & 50--99    & 0.0380  \\
Famous papers     & 100--499  & 0.0250  \\
Renowned papers   & 500+      & 0.00184 \\
\end{tabular}
\caption{The search option ``citation summary'' at the SPIRES
website returns the number of papers for a given author in the
categories in this table. The probabilities of getting citations
in these are intervals are listed in the third
column.}\label{tab:citsum}
\end{ruledtabular}
\end{table}
If a total of $M$ papers were drawn at random on the citation
distribution, the most probable result, $P_{\rm max}$, would
correspond to $m_i = M p_i$ papers in each bin.  The quantity
\begin{displaymath}
r  = - \log_{10} (P / P_{\rm max} ) \ ,
\end{displaymath}
is a useful measure of this probability which is relatively
independent of the number of bins chosen.  Since $r$ provides
completely objective information about the probability of drawing
a given citation record at random given knowledge of citation
patterns in that field, it is particularly well-suited for
comparisons between fields.  It is equally meaningful to calculate
$r$ for authors who publish in several fields. The leap from the
improbability of a given author's citation record to conclusions
regarding author quality requires certain assumptions which cannot
be tested.  For example, to compare citation records in the
instrumentation category with those in the remainder of our data
set, it is necessary to make some a priori assumption about the
relative intrinsic quality of the two data sets.  While the
``democratic'' assumption of equal intrinsic quality is easiest,
it may or may not be accurate.  (In a Bayesian sense, it is
necessary to establish a prior distribution.)

Consider the following two authors in the SPIRES data base. Author
A has a total of 200 publications with 17, 70, 82, 23, 8, and 0
publications in each of the bins above and an average of 26
citations per article.  Author B has a total of 176 publications
with 18, 79, 57, 10, 9, and 3 publications in each bin and an
average of 46. A simple calculation reveals that $r = 18.4$ for
Author A and $9.9$ for Author B.  The minimum value of $r$ is
evidently $0$. The maximum value of $r$ in the current data set is
found for Author C, who has a total of 217 publications with 5,
14, 38, 30, 97, and 33 publications in each of the bins above and
an average of 259 citations per article.  This leads to vastly
improbable value of $r = 181.3$.  With a total of 56224 citations,
Author C accounts for more than 1.5\% of all citations in the data
set.  There are also indications of less favorable correlations.
Author D has a total of 41 publications with 18, 23, 0, 0, 0, and
0 in each of the bins above and an average of $<1$ citation per
article.  This resulting value of $r = 4.43$ underscores the fact
that an improbable citation record is not necessarily a ``good''
one.

Given the total population of authors in SPIRES, these numbers
offer an objective indication of the extreme improbability that
the citation records of Authors A, B, and C were drawn at random.
These examples are far from exceptional. There are strong
correlations in the citation data, and they merit quantitative
study.  The differences between the Authors A and B can appear
surprising at first glance and emphasize the importance of a
priori criteria.  Although Author B has an average citation rate
almost twice that of Author A, his citation record is {\em more\/}
probable by a factor of $10^8$.  This is a natural consequence of
the power law distribution which makes it far more improbable to
have 10 papers with 100 citations each than one paper with 1000
citations.  The question of which of these options is ``better''
requires a subjective answer, and it is unlikely that any single
quantitative measure will satisfy everyone.  Thus, although the
interpretation of non-statistical fluctuations in individual
citation records is subjective, the likely presence of such
fluctuations can be identified with ease and objectivity.

It is as easy to calculate the $r$ for departments as for
individual authors.  Physics Department $\Delta$, which includes
Author C, published a total of 1309 papers from 1980 to 2000,
distributed with 81, 324, 474, 175, 216, 39 papers in the citation
summary bins. This results in a $r = 285$.  Physics Department
$\Gamma$, which includes Authors A and B, published a total of
1309 papers during the same period with $81$, $388$, $378$, $77$,
$28$, $3$. This yields the somewhat smaller value of $r = 65.9$.
Such information can be of practical value since it is seems
likely that the ``most improbable'' departments will have the
greatest success in attracting the ``most improbable'' authors.

\section{Summary and Conclusions}

We have considered the citation distribution of for 257,022
articles in the SPIRES data base and demonstrated the homogeneity
of topologies in the categories of theory, experiment and
phenomenology. The resulting data set can be reasonably described
by a simple power law with different exponents in the low and high
citation regions or a stretched exponential. Power law behavior is
a trait that the SPIRES data base shares with many other real
world networks, most notably the www
\cite{albert:99,kumar:99,broder:99}. It is clear that the
structures of these two networks are similar in many ways, with
scientific papers corresponding to \texttt{.html} documents. There
are differences, however. For example, because scientific papers
are printed, links are rarely bi-directional (reciprocity $\approx
0$); this is not the case for the www, where a non-vanishing
fraction of web-pages are bi-directional in spite of the directed
nature of hyperlinks.

The most striking features of the data include the extremely large
number of minimally cited papers and the fact that a remarkably
small number of papers (4\%) account for half of the citations in
the data set. While it is a truism that progress in physics is
driven by a few great minds, it can be disturbing to confront this
quantitatively.  The picture which emerges is thus a small number
of interesting and significant papers swimming in a sea of
``dead'' papers.  This has the practical consequence that any
study seeking to understand the dynamics of interesting papers
will be forced to discard most papers and accept the greatly
increased statistical uncertainties.  In the case of the SPIRES
data set, this would amount to roughly 10,000 papers.

In fact, the situation is even more dramatic due to the strong
correlations in the data set when considered as a function of
individual authors or individual institutions.  As we have seen in
the case of Author C above, a single author accounts for more than
1.5\% of all citations in the SPIRES data set.  Seven authors, not
necessarily the highest cited, account for 6\% of all the
citations.  We have suggested the measure of ``unlikelihood'',
$r$, defined above as a useful indicator of the presence of such
correlations. Further, this measure offers a tool for comparing
citation records in different fields with a known and controllable
bias.  (Any comparison across field boundaries must necessarily
involve unsupported assumptions and biases.  It is best to make
such assumptions visible and to discuss them.)  It would be
extremely valuable to perform ``longitudinal'' studies of citation
data collected as separate events. (An ``event'' here would be the
citation record of a single individual or single institution.)
This would permit a far more systematic study of the nature of the
statistically independent correlations and the probabilities with
which they occur. These strong correlations in the network
separates this particular network from many other small-world
networks, and constitute yet another difference between the www
and the network of scientific citations.

We emphasize that no single measure, such as our $r$ or the more
traditional average number of citations per paper, can claim to
capture the richness of either the full citation data set or
individual citation records.  While this is obvious from the
presence of strong correlations in the data, it is also supported
by the dramatic difference between the mean and median number of
citations in the global distributions reported here.  For this
reason, we believe that the value of large data bases such as
SPIRES and ISI would be greatly enhanced if global citation
distributions, such as that given in Figure 3 above, were
collected by subfield and made available to the users of these
data bases.

\begin{acknowledgments}
The authors would like to thank the helpful staff at the SPIRES
data base and its DESY and Durham mirrors, especially Heath
O'Connell and Travis C. Brooks at SLAC and Mike Whalley at Durham
University.
\end{acknowledgments}

\bibliography{speciale}
\bibliographystyle{unsrt}

\end{document}